\documentclass{article}
\usepackage{fullpage}
\usepackage{times}
\usepackage{amssymb}
\usepackage{graphicx}
\usepackage{amsmath}
\usepackage{amsthm}
\usepackage{hyperref}
\usepackage{subfigure}
\usepackage{bigints}

\input{tcilatex}
\begin{document}

\title{New Singularities in Unexpected Places}
\author{John D. Barrow\thanks{%
Email: J.D.Barrow@damtp.cam.ac.uk}\ {}\ and Alexander A. H. Graham\thanks{%
Email: A.A.H.Graham@damtp.cam.ac.uk} \\
%EndAName
\emph{Department of Applied Mathematics and Theoretical Physics}\\
\emph{University of Cambridge}\\
\emph{Wilberforce Road, Cambridge, CB3 0WA, UK}}
\date{\today}
\maketitle

\begin{abstract}
Spacetime singularities have been discovered which are physically much
weaker than those predicted by the classical singularity theorems. Geodesics
evolve through them and they only display infinities in the derivatives of
their curvature invariants. So far, these singularities have appeared to
require rather exotic and unphysical matter for their occurrence. Here we
show that a large class of singularities of this form can be found in a
simple Friedmann cosmology containing only a scalar-field with a power-law
self-interaction potential. Their existence challenges several preconceived
ideas about the nature of spacetime singularities and impacts upon the end
of inflation in the early universe.

\textbf{\ }
\end{abstract}

A striking feature of relativistic cosmology is the prediction that past and
future singularities can occur. Originally, singularities were defined by
the existence of incomplete geodesics, and a variety of sufficient
conditions for geodesic incompleteness were established by a series of
important theorems from 1965-1972 \cite{HE}. More recently, by using the
Einstein equations, new types of physical singularities have been identified
which can occur at finite time and are unaccompanied by geodesic
incompleteness \cite{barrow86,barrow04,barrow04a}. Many quantities, such as
the density and the expansion rate, which diverge at traditional 'big bang'
singularities, remain finite whilst other physical quantities, like the
pressure, diverge in finite proper time. The simplest example of what is
termed a 'sudden' singularity occurs in the zero-curvature Friedmann
universe with scale factor $a(t)$ and Hubble rate $H=\dot{a}/a$, containing
matter with density $\rho $ and pressure $p$. The field equations are ($8\pi
G=1=c$)

\begin{eqnarray}
3H^{2} &=&\rho ,  \label{1} \\
\dot{\rho} &=&-3H(\rho +p),  \label{2} \\
\ddot{a} &=&-\frac{(\rho +3p)a}{6},  \label{3}
\end{eqnarray}%
These equations permit there to be a finite time, $t_{s}$, at which $a,H,$
and $\rho $ all remain finite, in accord with Eq. (\ref{1}), but where $p,%
\dot{\rho}$ and $\ddot{a}$ all become infinite, in accord with Eqs. (\ref{2}%
)-(\ref{3}). The key to their existence is in not assuming any functional
link between $p$ and $\rho ,$ nor any boundedness condition on $p$, and this
freedom allows an acceleration singularity $\ddot{a}\rightarrow \infty $ to
arise at finite time as $t\rightarrow t_{s}$ because of a divergence in the
matter pressure, $p\rightarrow \infty $. Here is an explicit example. On the
time interval $0\leq {}t\leq {}t_{s},$ we can choose a solution for the
scale factor $a(t)$ of the form \emph{\ } 
\begin{equation}
a(t)=\left( \frac{t}{t_{s}}\right) ^{q}\left( a_{s}-1\right) +1-\left( 1-%
\frac{t}{t_{s}}\right) ^{n},  \label{A}
\end{equation}%
where $a_{s}\equiv a(t_{s})$, $q$ and $n$ are positive constants. If $%
t\rightarrow t_{s}$ from below then $a\rightarrow a_{s}$, $H\rightarrow
H_{s} $ and $\rho \rightarrow \rho _{s}>0,$ where $a_{s},H_{s},$ and $\rho
_{s}$ are all finite, but $p\rightarrow \infty $ and $\ddot{a}\rightarrow
-\infty $ whenever $1<n<2$ and $0<q\leq 1$. As $t\rightarrow 0$ we have a
big bang singularity with $\rho \rightarrow \infty $ and $a(t)\propto t^{q}$
but, as $t\rightarrow t_{s}$ is approached from below, a sudden singularity
occurs with $\ddot{a}\rightarrow -\infty $ but $a$ and $\dot{a}$ finite.

%\begin{equation}
%a(t)\rightarrow a_{s}+q(1-a_{s})\left(1-\frac{t}{t_{s}}\right)\rightarrow a_{s}.
%\label{asy}
%\end{equation}%
Nothing singular happens to geodesics as $t\rightarrow t_{s}$ \cite{fern04}
and we always have $\rho +3p>0$ because $\ddot{a}/a<0$. These singularities
are notable because they obey all the classical energy conditions bar the
dominant energy condition, in contrast to most other exotic singularities
discovered so far \cite{cald}. We can also create a divergence in any higher 
$(N+1)^{st}$ order time-derivative of the scale factor, with all lower-order
derivatives staying finite, by choosing $n\in (N,N+1)$ for integer $N>2$ in
Eq. (4). Adding the curvature term to the Friedmann equation makes no
difference to these conclusions, and Eq. (4) is actually a leading-order
approximation to part of the general solution of the Einstein equations \cite%
{barrow10}.

Notice that no equation of state linking $\rho $ and $p$ has been assumed,
and in fact the relation between them in these solutions tends to be
pathological, as $P$ diverges at finite $\rho$. It is natural to ask whether
this type of finite-time singularity can occur when there is a physically
motivated choice of $p$ and $\rho $ which doesn't allow them to be
independent variables? We shall show that the answer to this question is
'yes'.

There are in fact very simple examples that can arise in the study of
inflation in the early stages of the universe, or the universe's more recent
phase of dark-energy driven acceleration. They arise for the simple case of
a cosmological scalar field, $\phi $, with a positive power-law
self-interaction potential, $V(\phi )$, contributing a density and pressure 
\begin{equation*}
\rho =\frac{1}{2}\dot{\phi}^{2}+V(\phi )\ \text{ and }\ p=\frac{1}{2}\dot{%
\phi}^{2}-V(\phi ),
\end{equation*}
with 
\begin{equation}  \label{5}
V(\phi )=A\phi ^{n},\text{ }A>0.
\end{equation}

The field equations (\ref{1})-(\ref{3}) are therefore%
\begin{eqnarray}
3H^{2} &=&\frac{1}{2}\dot{\phi}^{2}+V(\phi ),  \label{f1} \\
\ddot{\phi}&=&-3H\dot{\phi}-An\phi^{n-1} ,\text{ }  \label{f2} \\
\text{ }2\dot{H} &=&-\dot{\phi}^{2}.  \label{f3}
\end{eqnarray}%
When $n$ is a positive \textit{even} integer they describe the classic model
of large-field inflation in a potential with a single minimum \cite{linde83}%
. When $n$ is a positive \textit{odd} integer the universe appears to
recollapse under the influence of the scalar field (for the $n=1$ case see
Ref. \cite{pad}). We will be interested in the case where $n>0,$ with $n$ 
\textit{not} an integer.

We first examine the case where $0<n<1$. We choose initial conditions so
that the universe is expanding initially and $\phi _{0}>0$, but the value of 
$\dot{\phi}_{0}$ is unconstrained. It is not difficult to see how the system
evolves in time. Since $\dot{\phi}_{0}>0$ both terms on the right-hand side
of Eq. \eqref{f2} are negative, so in finite time $\dot{\phi}$ becomes
negative. Hence, in finite time the scalar field starts to decrease, and
since $\dot{\phi}$ continues to decrease, as the second term on the
right-hand side of Eq. \eqref{f2} increases as $\phi $ decreases, it will
reach $\phi =0$ in finite time. When this happens $\dot{\phi}$ can also be
shown to be finite and strictly negative using Eq. \eqref{f1}, but from Eq. %
\eqref{f2} we have that $\ddot{\phi}\rightarrow -\infty $ as $\phi
\rightarrow 0$. From Eqs. \eqref{f1}-\eqref{f3}, we see that $H$ and $\dot{H}
$ are both finite at this point but $\ddot{H}$ diverges because 
\begin{equation}
\ddot{H}=-\dot{\phi}\ddot{\phi}\rightarrow -\infty \ \mbox{ as }\ \phi
\rightarrow 0.  \label{7}
\end{equation}%
This divergence is not a scalar polynomial curvature singularity \cite{ES},
as both $H$ and $\dot{H}$ are finite at this point. For our spatially-flat
Friedmann universe, the Ricci scalar, $R$, may be written as 
\begin{equation}
R=6(2H^{2}+\dot{H}),  \label{8}
\end{equation}%
which is clearly finite as $\phi \rightarrow 0$. However, higher scalar
derivatives of the curvature (like $\partial _{a}R\partial ^{a}R$ or $\Box
{}R$) are not regular since 
\begin{equation}
\dot{R}=6(4H\dot{H}+\ddot{H})\rightarrow -\infty \ \mbox{ as }\ \phi
\rightarrow 0.  \label{9}
\end{equation}
It is easy to check also that these singularities satisfy all the classical
energy conditions in the vicinity of the singularity. This is the first
example of a finite-time singularity for a simple and realistic matter model.

Similar singularities can also be shown to exist when $n>1$, although in
this case only higher-order derivatives of $\phi $ will diverge at the
singularity. This may be seen as differentiating Eq. (\ref{f2}) once gives 
\begin{equation}
\dddot{\phi}-9H^{2}\dot{\phi}-\frac{3}{2}\dot{\phi}^{3}-3HV^{\prime }(\phi
)+V^{\prime \prime }(\phi )\dot{\phi}=0.  \label{15}
\end{equation}%
For $1<n<2,$ every term except the first and last on the left-hand side is
finite as $\phi \rightarrow 0$, so $\dddot{\phi}\rightarrow \infty $ as $%
\phi \rightarrow 0$. This means that the first divergence in the scale
factor occurs at fourth order in its derivatives, since 
\begin{equation}
\dddot{H}=-\ddot{\phi}^{2}-\dot{\phi}\dddot{\phi}\rightarrow \infty \ 
\mbox{
as }\ \phi \rightarrow 0.  \label{16}
\end{equation}%
Hence, $\Box {}R$ and higher derivatives of the curvature are divergent on
approach to this singularity.

It is not difficult to generalise these conclusions to scalar-field
potentials of the form of Eq. (4) with arbitrarily large non-integer values
of $n$. If $k<n<k+1$, where $k$ is a positive integer, then as $\phi
\rightarrow 0$ we have $\phi ^{(k+2)}\rightarrow (-1)^{k+1}\times \infty $,
with all lower derivatives of $\phi $ finite. This implies that the first
divergence of the Hubble rate occurs for the $(k+2)^{th}$ derivative: $%
H^{(k+2)}\rightarrow (-1)^{k+1}\times \infty $ as $\phi \rightarrow 0$. But
if $n$ is an integer these singularities \textit{never} occur because $%
V(\phi )$ is smooth at $\phi =0$. By similar arguments any potential which
is not smooth at $\phi =0$ should create singularities of a similar type.

These singularities have a number of remarkable properties. They are
remarkably weak in that they exhibit no divergence of the curvature on
approach to the singularity and all polynomial curvature invariants are
finite: the only divergence occurs in derivatives of the curvature. Due to
the weakness of these singularities the spacetime remains geodesically
complete. To our knowledge they are the first examples of such weak
singularities in a Friedmann spacetime with a realistic matter model.

Evolving these spacetimes beyond $\phi =0$ is not always simple, as in some
cases (for instance $n=\frac{1}{2}$) the matter model breaks down beyond the
singular point $\phi =0$, since the naive evolution would push $\phi $ to
become strictly negative, which makes the expansion rate become complex.
However, this is only the case for some choices of $n$, and there are many
choices (e.g. $n=\frac{1}{3}$) for which $V(\phi )$ is always real-valued.
Numerical evidence suggests that if $V(\phi )$ is negative-definite for $%
\phi <0$ the spacetime collapses to a 'big crunch' singularity, while if $%
V(\phi )$ is positive-definite for $\phi <0$ no such collapse occurs.
Instead, the universe passes through $\phi =0$ an infinite number of times.
These spacetimes illustrate that the distinction between singular and
non-singular spacetimes is by no means clear-cut, and there are many
spacetimes, sourced by realistic matter, which are geodesically complete yet
possess observables which can evolve to finite-time singularities.

The formation of these singularities is completely generic. Indeed, once can
rigorously show that they form from any homogeneous and isotropic initial
data when $0<n<1$. They can also be shown to be stable to small
perturbations using standard perturbation theory \cite{barrow09}, and one
can adapt the arguments above to show they form in expanding, homogeneous
universes with large anisotropies. They are even stable to quantum
corrections since one can construct examples where the divergence only
occurs at an arbitrarily high order of the scale factor derivative \cite%
{fab, nojiri04}. This is in contrast to most other examples of weak
singularities discovered so far.

Finally, the model studied in this essay offers an intriguing alternative to
conventional models of inflation. If we choose initial conditions so that
the system starts high enough up the potential, and its early evolution is
potential-dominated, then inflation occurs as usual for as long as the
strong-energy condition is violated. Inflation ends as $\phi {}\rightarrow
0, $ whereupon the universe enters the reheating phase \cite{kofman94}.
However, the evolution will differs from standard models of inflation when
the system reaches $\phi =0$ deep in the reheating phase. When that happens,
the spacetime develops the weak singularity described in this essay. Since
predictions for the power spectrum of the cosmic microwave background (CMB)
are insensitive to the behaviour at reheating these models will give the
same predictions for CMB observables as conventional large-field inflation
models. Indeed, our monomial potentials with $n<2$ give a better fit to
current CMB data than those with large integer values of $n$ \cite{planck}.
However, these models ultimately have very different dynamics from
conventional reheating models. They bring inflation to a singular but timely
end.

%\bibliography{bh1ref}

\begin{thebibliography}{99}
\bibitem{HE} S.W. Hawking and G.F.R. Ellis, \textit{The large scale
structure of space-time}, (CUP, Cambridge, 1973)

\bibitem{barrow86} J. D. Barrow, G. J. Galloway and F. J. Tipler, \emph{Mon.
Not. Roy. Astron. Soc.} \textbf{223}, 835-844 (1986)

\bibitem{barrow04} J. D. Barrow, \emph{Class. Quantum Grav.} \textbf{21},
L79-L82 (2004)

\bibitem{barrow04a} J. D. Barrow, \emph{Class. Quantum Grav.} \textbf{21},
5619-5622 (2004)

\bibitem{fern04} L. Fern\'{a}ndez-Jambrina and R. Lazkoz, \emph{Phys. Rev. D}
\textbf{70}, 121503(R) (2004)

\bibitem{cald} R. R. Caldwell, \emph{Phys. Lett. B} \textbf{545}, 23 (2002)

\bibitem{barrow10} J. D. Barrow, S. Cotsakis and A. Tsokaros, \emph{Class.
Quantum Grav.} \textbf{27}, 165017 (2010)

\bibitem{linde83} A. D. Linde, \emph{Phys. Lett. B} \textbf{129}, 177-181
(1983)

\bibitem{pad} N. Kaloper and A. Padilla, \emph{Phys. Rev. Lett.} \textbf{114}%
, 101302 (2015)

\bibitem{ES} G. F. R. Ellis and B. Schmidt, \emph{Gen. Rel. Grav.} \textbf{10%
}, 989-997 (1979)

\bibitem{barrow09} J. D. Barrow and S. Z. W. Lip, \emph{Phys. Rev. D} 
\textbf{80}, 043518 (2009)

\bibitem{fab} J. D. Barrow, A. B. Batista, J. C. Fabris and S. Houndjo, 
\emph{Phys. Rev. D} \textbf{78,} 123508 (2008)

\bibitem{nojiri04} S. Nojiri and S. D. Odintsov, \emph{Phys. Lett. B} 
\textbf{595}, 1-8 (2004) and \emph{Phys. Rev. D} \textbf{70}, 103522 (2004)

\bibitem{kofman94} L. Kofman, A. Linde and A. A. Starobinsky, \emph{Phys.
Rev. Lett.} \textbf{73}, 3195-3198 (1994)

\bibitem{planck} P. A. R. Ade et al, \emph{Astron. Astrophys.} \textbf{571},
A22 (2014)

%\bibitem{page} D. N. Page, \emph{Class.Quantum Grav.} \textbf{1}, 417-427
%(1984)
\end{thebibliography}

\end{document}